\documentclass[12pt]{iopart}
\pdfoutput=1
\usepackage{color}
\usepackage{amssymb}
\usepackage[normalem]{ulem}
\usepackage{graphicx,wasysym}

\def\beqra{\begin{eqnarray}}
\def\eeqra{\end{eqnarray}}
\def\beq{\begin{equation}}
\def\eeq{\end{equation}}

\def\bV0{{\bf{V_0}}}
\def\re#1{(\ref{#1})}

\def\bs{{\bf{s}}}

\begin{document}

\title{Extracting the BAO scale from BOSS DR12 dataset }

\author{Eugenio Noda$^{1,2}$, Marco Peloso$^{3,4}$, Massimo Pietroni$^{1,4}$}
\vskip 0.3 cm
\address{

$^1$Dipartimento di Scienze Matematiche, Fisiche ed Informatiche dell'Universit\`a di Parma, Parco Area delle Scienze 7/a, I-43124 Parma, Italy\\
$^2$INFN, Gruppo Collegato di Parma, Parco Area delle Scienze 7/a, I-43124 Parma, Italy\\
$^3$Dipartimento di Fisica e Astronomia ``G. Galilei", Universit\`a degli Studi di Padova, via Marzolo 8, I-35131 Padova, Italy\\ 
$^4$INFN, Sezione di Padova, via Marzolo 8, I-35131, Padova, Italy\\
}

\begin{abstract}
We present the first application to real data from the BOSS DR12 dataset of the Extractor procedure \cite{Noda:2017tfh,Nishimichi:2017gdq} to determine the acoustic scale imprinted on Baryonic Acoustic Oscillations (BAO). We show that, being largely insensitive to the broadband shape of the Power Spectrum, this procedure requires a lower number of nuisance parameters than those used by the BOSS collaboration, For non-reconstructed data our analysis improves the accuracy on the acoustic scale by about 20 \%, while for reconstructed ones we get  essentially the same  level of accuracy as the BOSS analysis.  
\end{abstract}

\maketitle

\section{Introduction}
Measuring the acoustic scale from BAO's is one of the main goals of present and forthcoming galaxy surveys. A sub-percent level measurement will give powerful constraints on the nature of Dark Energy and on possible modifications of General Relativity.


In practical data analyses, the nonlinear effects, which include redshift space distortions (RSD) and bias, are modelled by adding a number of nuisance parameters. In particular, in the BOSS analysis that we will use as a benchmark in this paper, namely that based on the DR12 dataset presented in \cite{Beutler:2016ixs}, 10  parameters where introduced, of which only one was related to the observable of interest, the BAO scale, while the role of most of the remaining ones is to model UV effects.  However, it is well known \cite{Eisenstein:2006nj} that the information on the BAO scale is, to a great extent, insensitive to short scale effects, and can be robustly extracted from data. In particular, in \cite{Noda:2017tfh,Nishimichi:2017gdq} we introduced an ``Extractor'' procedure, which, given a Power Spectrum (PS), projects out the smooth component and gives only the oscillating one. The leading nonlinear effect on BAO wiggles is the well understood damping caused by random motions coherent on large (O$(10-100\; \rm{h^{-1}\,Mpc}))$ scales. Once this effect is taken into account, the `extracted' PS can be accurately modelled by as few as one nuisance parameter, which combines RSD and scale-dependent bias. The Extractor procedure was tested in \cite{Nishimichi:2017gdq} using N-body simulations as data and different models, with and without the inclusion of UV effects.

In this paper we test the Extractor procedure on real data, namely on  the already mentioned BOSS DR12 dataset \cite{Beutler:2016ixs}. Our main purpose is to assess the improvement induced by the strong reduction in the number of nuisance parameters allowed by our procedure. Therefore, we will stick to a very simple model for the PS, given in Eq.~\re{eq:model_BOSS}, which, from a computational point of view, requires no more that common 1-loop integrals in SPT. We will not consider more elaborate theoretical models for the PS, like the TRG+UV effects of \cite{Noda:2017tfh} or the Effective Field Theory of \cite{Carrasco:2012cv}, although they can be included in the analysis with no extra effort.

Finally, we stress that our analysis, at this  stage, focusses on the extraction of an oscillatory feature of a PS, rather than on the more general question of how to estimate cosmological distances in a model independent way (for a recent discussion, see for instance \cite{Anselmi:2018vjz}). Therefore, to be concrete, and to make a direct comparison with the results of \cite{Beutler:2016ixs} we concentrate on the $\alpha$ dilation parameter introduced in Sect.~ \ref{sec:BAOdetermine}. 

The paper is organized as follows. In Sect.~\ref{definition} we recall the definition of the BAO Extractor procedure, in Sect.~\ref{sec:data_reconstr} we summarize the aspects of the DR12 dataset which are more relevant to the following, in Sect.~\ref{sec:model} we introduce our model for the galaxy PS, in Sect.~\ref{sec:BAOdetermine} we describe our analysis, in Sect.~\ref{sec:windows} we report the convolutions necessary to account for the BOSS observation window functions,  in Sect.~\ref{smock} we test our procedure on the catalogue of mock galaxies used by the BOSS collaboration, and in Sect.~\ref{sdata} we perform the analysis on the real data. Finally, in Sect.~\ref{sconclu} we discuss our results and conclude.

\section{BAO extractor: definition}
\label{definition}

In this Section we summarise the procedure to extract the oscillatory part of a given PS, $P(k)$, that we use in this paper \cite{Noda:2017tfh}. As this procedure was already summarized in \cite{Nishimichi:2017gdq}, we only present a very short description here, referring the interested reader to that discussion for more details. 

The PS of Large Scale Structure (LSS) can be seen as the sum of a dominant smooth (``no-wiggle'') part and of an oscillatory (``wiggly'') component due to BAO: 
\begin{equation}
P(k) = P^{nw} (k) + P^w (k) \;\;,\;\; P^w(k) \simeq  P^{nw} (k)A(k) \, \sin \left( k r_{\rm bao} \right) \;, 
\label{e1}
\end{equation}
where the smooth function $A(k)$ damps the BAO beyond the Silk scale, and $r_{\rm bao}$ is the scale of the BAO  in the data. To extract the oscillatory part of the PS, we perform the operation  \cite{Noda:2017tfh}

\begin{equation}
R[P] \left( k; \Delta \right) \equiv  \frac{\frac{1}{2\, \Delta\, k_s} \int_{- \Delta\,k_s}^{ \Delta\,k_s}\left(1-\frac{P(k-q)}{P(k)}\right)  \,dq}{1-\frac{\sin(2\pi\Delta) }{2\pi\Delta}}\;, 
\label{R[P]} 
\end{equation} 
where $k_s$ is a wavenumber of order of the BAO scale. It only sets the scale of the integration range in $q$, and does not need to coincide with the actual scale of the oscillations contained in the PS, which we indicate with $r_{\mathrm{bao}}$. The integral at the numerator of (\ref{R[P]}) projects out the smooth broadband part of the PS and is insensitive to any constant linear bias, since it contains ratios of PS's. In  \cite{Nishimichi:2017gdq}, we made use of N-body simulations to show that the output of this extracting procedure can be reliably modelled as to include the effects of the nonlinear evolution of the matter field, of redshift space distortions and of scale-dependent halo bias (allowing to reproduce these effects with sub-percent accuracy). In particular, we showed that the effect of short scale (``UV'') physics on the extracted PS is very mild, as it mostly modifies the broadband (smooth) shape of $P^{nw} (k)$.

We also verified that the results obtained from the Extractor are very weakly dependent on the parameter $\Delta$, which sets the integration range in $q$, in units of $k_s$.  It should only be $O(1)$, otherwise, for $\Delta \gg 1$, the integration would smoothen out the oscillatory feature, while, for $\Delta \ll 1$ the advantage of using ocillatory information from nearby bins would be lost. In this paper we set $\Delta=0.6$. 

For a discrete set of data $\left\{ k_i ,\, P_i = P \left( k_i \right) \right\}$, the relation (\ref{R[P]}) can be written as 
\begin{equation}
R_i \equiv R \left[ P \right] \left( k_i \right) = D_i^{-1} \sum_l f_{i,l} \left( 1 - \frac{P_{l}}{P_i} \right) \;, 
\end{equation}
where
\beq
f_{i,l} = \Bigg\{
\begin{array}{l c}
1 & \;\;\;\; \textrm{if $l$ is such that $|k_l - k_i | \leq \Delta k_s $, and $i\neq l$}  \\
0 & \textrm{otherwise}
\label{def-fij}
\end{array}
\eeq
and
\beq
D_i= \sum_l f_{i,l} \left\{ 1- \cos \left[r_s\left(k_l - k_i\right) \right] \right\}.
\label{def-Di}
\eeq

To use the extractor to obtain informations on a given cosmology, we compare the extractor $R[P_{\rm data}]$ obtained from the data with that  $R[P_{\rm model}]$ of a given theoretical model. Specifically, in this work we compare the BOSS data  \cite{Beutler:2016ixs} with the model PS introduced in Section \ref{sec:model}. 

\section{The BOSS data and reconstruction} \label{sec:data_reconstr}
The Baryon Oscillation Spectroscopic Survey (BOSS) was part of the effort of the SDSS-III collaboration to map  our near Universe. It measured the spectroscopic redshift of luminous galaxies in the $0.2< z < 0.75$ redshift range and two regions of the sky, denoted by North Galactic Cup (NGC) and South Galactic Cup (SGC), from which, respectively, about 865\ 000 and 330\ 000 galaxies were measured \cite{Alam:2016hwk}. The measurements were made with a 2.5 metre-aperture telescope at the Apache Point Observatory, New Mexico, and covered about 10 000 square degrees. The data is publicly available and accessible through the SDSS-III website \cite{BOSSweb}. 

In our analysis we are interested in the PS, of which the BOSS collaboration measured the monopole, quadrupole, and hexadecapole moments. These were divided into 3 overlapping redshift bins: $0.2 < z < 0.5$, $0.4 < z < 0.6 $ and $0.5 < z <0.75$, for which the effective redshifts of $z_{eff} =$ 0.38, 0.51 and 0.61 were used in our computations.
The experimental data is accompanied by a set of mock galaxies, the MultiDark-Patchy mock catalogues \cite{Kitaura:2013cwa, Kitaura:2015uqa} whose primary purpose is to provide the covariance matrices, but can also be used to test the analysis pipeline.

Nonlinear dynamics damps the BAO signal which, if not properly accounted for, degrades to a large extent the information that can be extracted from the data. One approach is to properly model the damping using the techniques we have presented in \cite{Noda:2017tfh, Nishimichi:2017gdq}. Another approach is to act on the data itself using so called \textit{reconstrution} techniques \cite{Eisenstein:2006nk}.
The aim of these techniques is to (partially) undo the bulk motions of galaxies, by moving them  back  to their original position, by reconstructing the velocity field from the density one, via the continuity equation.
In linear perturbation theory the  displacement $\Psi$  is related to the redshift-space density field by \cite{Nusser:1993sx}
\beq \label{eq:lagr_rec}
\nabla \cdot \Psi(\bs) + \beta \nabla \cdot (\Psi \cdot \hat{s}_{||}) \hat{s}_{||} = - \frac{\tilde{\delta}_{obs}(\bs)}{b}\,,
\eeq
where $b$ is the linear galaxy bias, $\hat{s}_{||}$ is the unit-vector along the line of sight, and $\beta = b/f$  (where $f$  is the linear growth function). The density field $\tilde{\delta}_{obs}(\bs)$ is  smoothed on a scale, which in the BOSS implementation of reconstruction \cite{Beutler:2016ixs} (originally implemented for data analysis in \cite{Padmanabhan:2012hf})  is taken to be $\Sigma_{sm} = 15 \,\rm{h}^{-1}  \textrm{Mpc}$ using a gaussian filter
\beq
S(k) = \exp\left[ -(k \Sigma_{sm})^2 /2 \right].
\eeq
Filtering the density to damp the nonlinearities at small scales is required in order to use of the linear continuity equation, Eq. \re{eq:lagr_rec}. 

Assuming the displacement field is irrotational, the equation above can be solved, and the the line-of-sight and angular positions of the galaxies are shifted as follows
\beqra
&&s_{||}^\mathrm{new}=s_{||}^\mathrm{old}-(1+f) \Psi_{||}(\bs^\mathrm{old})\,,\nonumber\\
&&s_{\perp}^\mathrm{new}=s_{\perp}^\mathrm{old}-\Psi_{\perp}(\bs^\mathrm{old})\,,
\label{reco}
\eeqra
where the $(1+f)$ factor multiplying the displacement along the line of sight aims at removing the linear component of redshift-space distortions. 

The same displacement procedure is applied to a set of randomly distributed particles. The reconstructed density field are then  given by the difference between the displaced galaxy field and the displaced random field.

Therefore we have, for each multipole and each redshift bin, 2 pre-reconstruction PS's (from the NGC and SGC) and 2 post-reconstruction PS's. In the following we are going to apply the extractor analysis to these sets of data and compare the results with those obtained by the analysis performed by the BOSS collaboration using a 10-parameter  fit for the model PS, both for the pre-reconstruction and post-reconstruction datasets.

\section{The model} \label{sec:model}

In \cite{Nishimichi:2017gdq} we tested the performance of a simple model function to reproduce the extracted PS obtained from N-Body simulations and found good agreement, both for matter and for halos, in real and redshift space. This is particularly interesting since the model did not include any short-scale effects, which are otherwise essential to model the full PS, and  both the scale dependence of RSD and halo bias are encoded in a single exponential prefactor containing just one extra parameter.

We will use a similar model to analyze BOSS experimental data, namely, 
\beqra \label{eq:model_BOSS}
\!\!\!\!\!\!\!\!\!\!\!\!\!\!\!\!\!\!\!\!\!\!\!\! P_{\rm{model}}(k,\mu;A,b) = e^{-A k^2}\Bigg[ \left(b+ \mu^2 f R_{sd}(k)\right)^2 \left(P^{nw, l}(k) + P^{w, l}(k) e^{-k^2 \Xi^{rs}(\mu)/\gamma_{rec} }\right) \nonumber \\
+ b^2 \Delta P_{\delta \delta}^{nw, 1l}(k) + 2 b \mu^2 f \Delta P_{\delta \theta}^{nw, 1l}(k) + \mu^4 f^2 \Delta P_{\theta \theta}^{nw, 1l} \Bigg] \,, 
\eeqra
where $\mu$ is the cosine of the angle between the wavevector and the line of sight, $P^{nw, l}$ and $P^{nw, l}$ are, respectively, the smooth and the oscillating components of the linear PS, while $\Delta P_{ij}^{nw, 1l}(k)$ ($i,j=\delta,\theta$) denote the components of the real space 1-loop PS, computed using the smooth linear one. The smooth PS is obtained by spline interpolating the linear PS on the points corresponding to nodes of $\sin(k\,r_\mathrm{bao})$, and the wiggly PS is then the difference between the linear PS and the smooth one.

The $R_{sd}(k)$ term accounts for the removal of large scale redshift space distortions by the reconstruction procedure discussed in the previous section, Eq.~\re{reco}, and in the BOSS analysis  the `Rec-Iso' convention of \cite{Seo:2015eyw} is adopted,
\beq
R_{sd}(k) = \left\{
\begin{array}{c l }
1 & \textrm{ \; before reconstruction},   \\
1- e^{- \frac{ k^2 \Sigma_{sm}^2}{2}} & \textrm{\;\, after reconstruction}\,. \\
\end{array} 
\right.
\label{damrs}
\eeq
The remaining, nonlinear, redshift space distortions effect is modelled, together with scale dependent bias, by the single pre-factor $e^{-A k^2}$. In general, this would be expected  to be a poor approximation, as, for instance, it  lacks any $\mu$-dependence. However, since, as we will see, we will only include monopole data in our BAO analysis, our simple $\mu$-independent parameterization can be considered as an effective one, after $\mu$-averaging. 

The effect of BAO damping by large-scale bulk flows is encoded in the $e^{-k^2 \Xi^{rs}(\mu)/\gamma_{rec} }$ term, where
the function $\Xi^{rs}(\mu)$ is given by \cite{Nishimichi:2017gdq}
\beqra
&&\Xi^{rs}(\mu;r_{bao})= \left(1+f\mu^2(2+f)\right)\Xi(r_{bao})\nonumber\\
&&\qquad\qquad\qquad\qquad +f\mu^2(\mu^2-1)\frac{1}{2\pi^2}\int dq\,P^{nw,l}(q) j_2(qr_{bao})\,,
\label{XiRS}
\eeqra
with
\beqra
&&
\Xi(r_{bao}) \equiv \frac{1}{6 \pi^2}\int  dq\,P^{nw,l}(q;z)\,( 1-j_0(q\,r_{bao}) + 2 j_2(q \,r_{bao}))\,,
\label{xi-bao}
\eeqra
with $j_n(x)$'s the spherical Bessel functions.
Typically, the reconstruction reduces the damping of the BAO of a factor $\sim 4$ \cite{Eisenstein:2006nj}. Hence we divide the exponential damping by a quantity $\gamma_{rec}$ which is equal to 1 for pre-reconstruction data and to 4 for reconstructed data.

From the PS model above we are going to calculate the multipole moments:
\beq
P_l(k;A,b) \equiv \frac{2l +1}{2} \int d\mu P_{\rm{model}}(k,\mu;A, b) {\cal{P}}_l(\mu),
\eeq
where ${\cal{P}}_l$ are the Legendre polynomials, and  $l=0,2$ correspond, respectively, to the monopole and quadrupole moment. 

We stress again that Eq.~\re{eq:model_BOSS}  does not include any short scale correction term as those implemented in the TRG approach discussed in \cite{Noda:2017tfh}, or in the Effective Field Theory of the LSS \cite{Carrasco:2012cv}. Here  we want to focus on the performance of the Extractor compared to the standard BAO analysis, as implemented by the BOSS collaboration, and our main goal is to assess the effect of the reduction in nuisance parameters present in their analysis. 
 
The BOSS analysis \cite{Beutler:2016ixs} of the monopole data is based on the power spectrum
\begin{equation}
P \left( k \right) = B^2 \, P_{\rm sm,{\rm lin}} \left( k \right) F_{\rm fog} \left( k, \Sigma_s \right) \, 
\left[ 1+ \left( \frac{P_{\rm lin} \left( k \right) }{P_{\rm sm,lin} \left( k \right)} - 1 \right) {\rm e}^{-k^2 \, \Sigma_{\rm nl}^2 / 2} \right] \;. 
\end{equation} 
In this expression, the factor $F_{\rm fog} \left( k, \mu, \Sigma_s \right) = \frac{1}{\left( 1+k^2 \, \Sigma_s^2/2 \right)^2}$ is a damping term due to the non-linear velocity field (``Finger-of-God''). The two power spectra $P_{\rm lin}$ and $P_{\rm sm,lin}$ are, respectively, the linear power spectrum of the assumed  cosmology, and a fit obtained by adding an Eisenstein $\&$ Hu \cite{Eisenstein:1997ik} no-Wiggle power spectrum, together with the five monomial terms \cite{Beutler:2016ixs} 
\begin{equation}
\frac{a_{0,1}}{k^3} +  \frac{a_{0,2}}{k^2} +  \frac{a_{0,3}}{k} + a_{0,4} + a_{0,5} \, k \;. 
\end{equation} 
The BOSS analysis is therefore characterized by $10$ parameters ($B^{\rm NGC}$, $B^{\rm SGC}$, $\alpha_{0,1-5}$, $\alpha$, $\Sigma_{\rm nl}$, $\Sigma_s$), to be compared to the $3$ parameters employed in the analysis presented in this paper ($A$, $b$, $\alpha$).

\section{Determining the BAO scale} \label{sec:BAOdetermine} 

Our aim is to determine the BAO scale contained in the data. To do this, following the BOSS analysis, we are going to define a parameter $\alpha$ that rescales isotropically the BAO length imprinted in the data with respect to our fiducial cosmological model.
In \cite{Nishimichi:2017gdq} we considered a function $\chi^2(\alpha)$,
\beq
\chi^2(\alpha,\cdots)=\sum_{i,j=n_{min}}^{n_{max}} \delta R[P](k_i;\alpha,\cdots) \;c_{ij}^{-1} \;\delta R[P](k_j;\alpha,\cdots) \,,
\eeq
with
\beq
\delta R[P](k_i;\alpha,\cdots)\equiv R[P_{model}](k_i/\alpha,\cdots) - R[P_{data}](k_i)\,,
\label{alpha-def}
\eeq
where $c_{ij}$ is the covariance matrix of Eq.~\re{cov-RR} below, and dots indicate other possible parameters  of the model. Notice the role of the $\alpha$ parameter in rescaling the  lenght scales: if the best fit is for $\alpha=1$ the model and the data agree on the BAO scale. 

In the present analysis, our model contains 2 nuisance parameters, $b$ and $A$. 
The parameter $b$ mainly fixes the overall normalization of the PS, to which the Extractor operator is essentially insensitive. 
The role of $A$ is  to account for  RSD and scale-dependent bias, and it mostly modifies the broadband shape of the PS, while  $\alpha$ carries information on the BAO scale imprinted on the wiggly component.


Therefore we will combine  constraints both from the broadband and the Extractor by considering a composite $\chi^2$ (in Eq. (\ref{chi2-sum}) below).
In doing so we need to take care of two aspects. Firstly, we need to consider the cross correlations between $P^0(k)$ and $R[P^0](k)$. Secondly, we must not use the same information twice, both in the full PS and the extracted one. In order to take care of these two issues we reformulate our procedure as follows:

First of all we define a list of indices:
\beq
\{ \pi_1, \dots , \pi_{max}, \rho_1, \dots, \rho_{max} \},
\eeq
where the $\pi_i$'s are the wavenumber bins used for the full PS, while the $\rho_i$'s are those used for the extracted PS. To make sure we avoid double counting the data, each bin is present at most once in the list, so it will not be used for the fit of the full and extracted PS at the same time.
We then define a vector:
\beqra
\textrm{X} \left( A,\, b ,\, \alpha \right) \equiv &\Big\{ \Delta_P \left( k_{\pi_1},\, A ,\, b  \right), \dots, \Delta_P \left( k_{\pi_{max}},\,A,\,  b  \right) , \nonumber \\
& \quad \quad \Delta_{RP} \left( k_{\rho_1}, A, b, \alpha \right), \dots, \Delta_{RP} \left( k_{\rho_{max}}, A,\, b ,\, \alpha \right)   \Big\},
\eeqra
where
\beq
\Delta_P \left( k_i,\, A ,\, b \right) \equiv P^0_{data} \left( k_i \right) - P^0_{model} \left( k_i,\, A ,\, b \right), 
\eeq
and 
\beq
\Delta_{RP} \left( k_i, \,A ,\, b ,\, \alpha \right) \equiv R[P^0_{data}](k_i) - R[P^0_{model}] \left( k_i/\alpha,\,  A ,\, b \right). 
\eeq

From these differences, we define the $\chi^2$ function as 
\beq
\chi^2 \left( A ,\, b ,\, \alpha \right) \equiv \sum_{i,j} \textrm{X}_i \left( A ,\, b ,\, \alpha \right) \, C^{-1}_{i,j} \, \textrm{X}_j \left( A,\, b ,\, \alpha \right), 
\label{chi2-sum}
\eeq
where we have used the covariance matrix elements 
\beq
C_{i,j} \equiv \left\{
\begin{array}{l r }
\langle \delta P_i \ \delta P_j \rangle & i \in (\pi_1,\pi_{max}), j \in (\pi_1,\pi_{max})  \\
\langle \delta P_i \ \delta R_{j} \rangle & i \in (\pi_1,\pi_{max}), j \in (\rho_1,\rho_{max}) \\
\langle \delta R_{i} \ \delta P_j \rangle & i \in (\rho_1,\rho_{max}), j \in (\pi_1,\pi_{max}) \\
\langle \delta R_{i} \ \delta R_{j} \rangle & i \in (\rho_1,\rho_{max}), j \in (\rho_1,\rho_{max}) \\
\end{array} \right.
\eeq
The first line of this expression contains the PS covariance matrix elements, $\langle \delta P_i \ \delta P_j \rangle $. 
These are obtained from the mock catalogue  \cite{BOSSweb}, that we discuss in the following section. The elements of the last three lines are given in terms of the former by the relations 
\beqra
\langle \delta R_i \delta P_j \rangle = D_i^{-1} P_i^{-2} \sum_l f_{i,l} \Big( - P_i \langle \delta P_l \ \delta P_j \rangle + P_l \langle \delta P_i \ \delta P_j \rangle \Big) 
\label{cov-RP}\,,
\eeqra
and
\begin{eqnarray} 
& & \!\!\!\!\!\!\!\!  \!\!\!\!\!\!\!\!  \!\!\!\!\!\!\!\!   \!\!\!\!\!\!\!\! 
\left\langle \delta R_i \, \delta R_j \right\rangle = D_i^{-1} P_i^{-2} D_j^{-1} P_j^{-2} \, \sum_{m,n} 
f_{i,m} f_{j,n} 
\Bigg[ P_m \, P_n \, \left\langle \delta P_i \delta P_j \right\rangle - P_m \, P_j \, \left\langle \delta P_i \delta P_n \right\rangle \nonumber\\ 
&& \!\!\!\!\!\!\!\!  \quad\quad  \quad\quad  \quad\quad  \quad\quad  \quad\quad 
 - P_i \, P_n \, \left\langle \delta P_m \delta P_j \right\rangle + P_i \, P_j \,  
\left\langle \delta P_m \delta P_n \right\rangle \Bigg] \;.
\label{cov-RR}
\end{eqnarray} 
We note that if we were to repeat one index in $\pi_i$ and $\rho_i$ the covariance matrix $C_{i,j}$ would become singular, making the analysis procedure fail. In fact, this feature works as a ``safety" mechanism that prevents us from inadvertently run into overfitting.

From the $\chi^2 \left( A ,\, b ,\, \alpha \right)$ function we construct a series of 1 dimensional probability distribution functions (PDF) for each redshift bin $z$, by combining the North and South samples, and by marginalizing over two parameters at a time. For instance, the PDF for the parameter $\alpha$ is obtained from 
\beq \label{eq:pdf}
f_{z}(\alpha) = \frac{1}{\cal{N}}\int dA \, d b \ e^{- \frac{\chi^2_{N,z} \left( A ,\, b ,\, \alpha \right)+ \chi^2_{S,z} \left( A ,\, b ,\, \alpha \right)}{2}} , 
\eeq
and analogously for the other two parameters. The factor ${\cal{N}}$ normalises the total probability to 1. As the following figures show, the PDFs for the parameter $\alpha$ present a peak that is very close to a Gaussian. For this reason, the following tables report the average and the standard deviation of $\alpha_i$, weighted by $f_{z} \left( \alpha_i \right)$, (where $\alpha_i$ are values on a discretized grid, where our analysis is performed) for the best estimate and the $68\%$ C.L. interval for this parameter. 

In constructing the PDF as shown above we are assuming that the parameters $A$ and $b$ are the same for the NGC and SGC samples. We have also tested the impact of assuming four separate parameters $A_N$,  $A_S$, $b_N$, $b_S$ (in addition to $\alpha$), and we have obtained values of $\chi^2$ that are only marginally better than those obtained from the common  $A$ and $b$ analysis; namely, the improvement in the $\chi^2$ does not justify including two extra parameters. For this reason, all our results below are obtained for $A_N = A_S \equiv A$, and for $b_N = b_S \equiv b$.

\section{Window function}
\label{sec:windows} 

To be compared to BOSS data, the model PS has to be convolved with the survey window function provided by the BOSS collaboration. The window-corrected PS is given by \cite{Beutler:2016ixs}
\beq
\hat P_l(k)=(-i)^l 4\pi\int dr \,r^2 j_l(k r) \hat\xi_l(r)\,,
\eeq
where the $\hat\xi_l(r)$'s are the window-corrected correlation function multipoles  (including only monopole and quadrupole)
\beqra
&&\hat\xi_0(r) = W_0^2(r)\,\xi_0(r)+\frac{1}{5}  W_2^2(r)\,\xi_2(r)\,,\nonumber\\
&&\hat\xi_2(r) = W_2^2(r)\,\xi_0(r)+\left(W_0^2(r)+\frac{2}{7}  W_2^2(r)\right) \,\xi_2(r)\,,
\eeqra
with $W_0^2(r)$ and $W_2^2(r)$ the monopole and quadrupole of the window function.
Expressing the correlation function in term of the PS,
\beq
\xi_l(r)=i^l\int\frac{dk k^2}{2 \pi^2} j_l(kr) P_l(k)\,,
\eeq
we can write the window-corrected monopole PS as
\beq
\hat P_0(k)= \int dq \left(A_0(k,q) P_0(q) -\frac{1}{5} A_2(k,q) P_2(q)\right)\,,
\eeq
where the convolution kernels are given by
\beqra
&& A_0(k,q)\equiv \frac{2 }{\pi} q^2 \int dr \,r^2\;j_0(k r) j_0(q r) W_0^2(r)\,,\nonumber\\
&& A_2(k,q)\equiv \frac{2 }{\pi} q^2 \int dr \,r^2\;j_0(k r) j_2(q r) W_0^2(r)\,.
\eeqra
An analogous expression holds for the window-corrected PS quadrupole, which, however, we will not include in our analysis.

\section{Test on mock galaxies}\label{smock}

\begin{table}
\centering
\begin{tabular}{|c|c|c|c|c|}
\hline
 & MDP Mock & Model 1 & Model 2 & Reference \\ \hline
$\Omega_m$  & $0.307115$ & 0.29 & 0.31 & 0.31 \\ \hline
$\Omega_b$  & $0.048206$ & 0.0458 & 0.0481 & 0.048 \\ \hline
$\sigma_8$  & $0.8288$ & 0.8493 & 0.8758 & 0.824 \\ \hline
$n_s$  & 0.9611 & 0.97 & 0.9624 & 0.96 \\ \hline
$h$  & 0.677 & 0.7 & 0.7 & $0.676$ \\ \hline
\end{tabular}
\caption{Cosmologies employed in this work. The first cosmology is the one used for the generation  of the MultyDark Patchy mock data, and for our analysis of these data presented in Table \ref{tab:results_mock}. The second and third cosmology are the ``wrong'' cosmologies employed for the test of the mock data presented in Table \ref{tab:results_wrong_cosm}. The final cosmology is the Reference Cosmology used by the BOSS analysis of ref. \cite{Beutler:2016ixs}, and for our analysis presented in Table \ref{tab:results_BOSS}. This cosmology also assumes nonvanishing neutrino masses $\sum m_\nu = 0.06 \, {\rm eV}$. 
}
\label{tab:cosmologies}
\end{table}

\begin{figure}
\includegraphics[width=\textwidth]{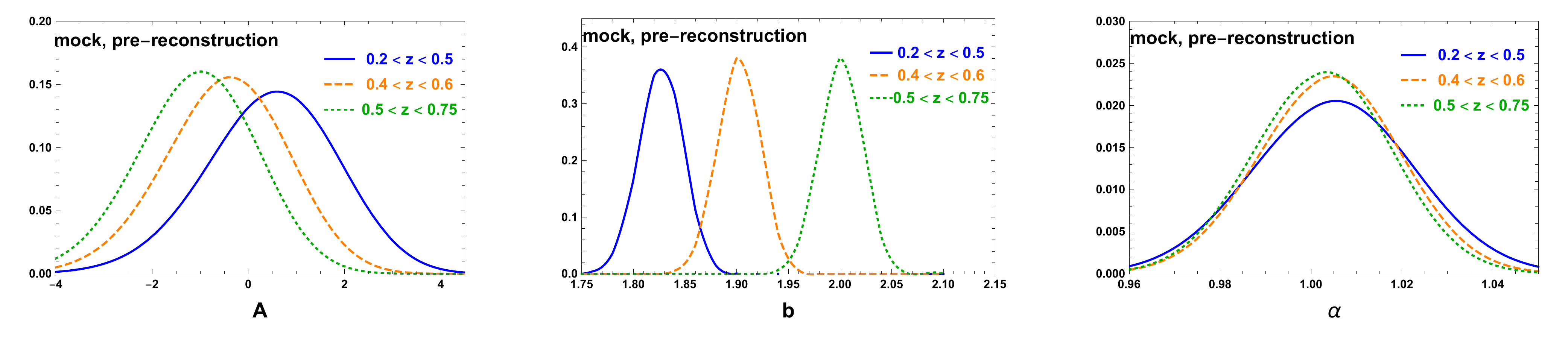} \\
\includegraphics[width=\textwidth]{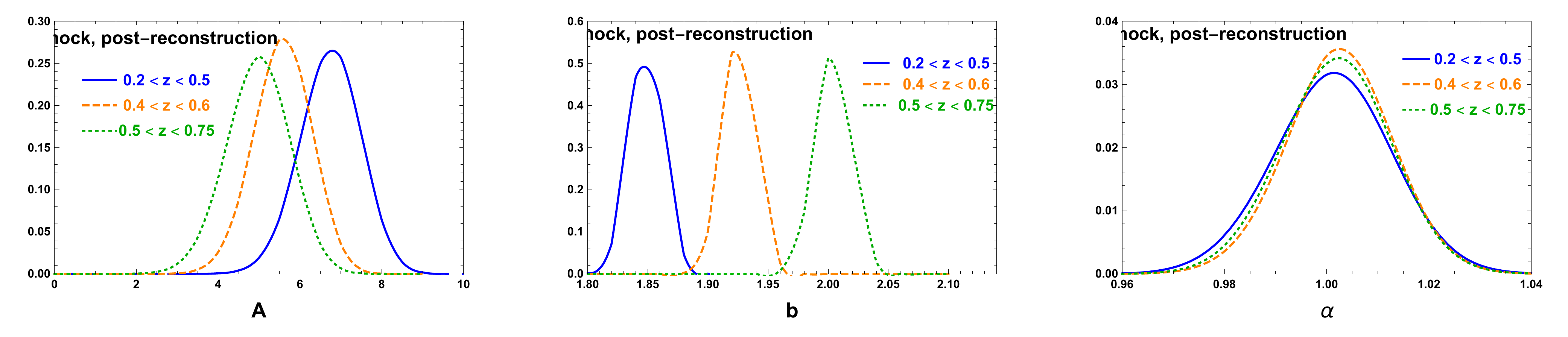} \\
\caption{Marginalized PDFs for the three parameters $A$, $b$, and $\alpha$, and for different redshift bins. 
The analysis refers to the mock galaxies, pre (first row) and post (scond row) reconstruction.}
\label{Fig:CHI2_mock}
\end{figure}

Before applying the procedure to the experimental data we tested it on the MultyDark Patchy mock catalogues of the BOSS collaboration \cite{Kitaura:2013cwa, Kitaura:2015uqa}. The fiducial cosmological model used to generate these mocks was a flat $\Lambda$CDM model with the parameters reported in the second column of Table \ref{tab:cosmologies}. The corresponding PDF for the three parameters and for the various redshift bins obtained from our analysis are shown in Figure~\ref{Fig:CHI2_mock}. 

Testing our analysis on the mock data has the main purpose of choosing the precise sets  $\pi$ and $\rho$ described in the previous section. We have compared a few different choices, and we have found that the following choices allow for a good description of the data (relatively lower $\chi^2$):   $\pi = \{1, 2,3,4 ,10, 12\}$, and $\rho=\{5,\dots,9,11,\, 13,\, \dots,22\}$ (respectively, $\rho=\{5,\dots,9,11,\, 13,\, \dots,23\}$) for the pre-recostruction (respectively, post-reconstruction) data. The index $i$ in these lists refer to the corresponding  wave number in the BOSS data (we recall that, approximately, the wavenumbers in the BOSS data are given by $k_i \simeq \left( 0.01 * i + 0.005 \right) \, {\rm h \, Mpc}^{-1}$). 

The optimal configuration is to use most bins in $\rho$, given that the extracted PS has greater constraining power on $\alpha$, and use just a few bins in $\pi$, that improve the constrains on $A$, and, particularly, on $b$.  Since the extractor is an integrated quantity, the first bins of R[P] are affected by boundary effects, so a natural choice is for them to be in $\pi$. Moreover, the BAOs can be neglected at such low wavevectors.  We then added two values ($i=10,12$) at moderate $k$, where the broadband part of the model power spectrum (\ref{eq:model_BOSS}) is still valid, so to have a larger ``arm'' for costraining $A$ and $b$. These two points cannot be also simultaneously used in the extractor part of the analysis, as so we chose points close to extrema of the BAO wiggles, which are least sensitive to the parameter $\alpha$ (since $\alpha$ is a shift in the horizontal axis, it mostly affects regions where $\left\vert \frac{d R \left[ P \left( k \right) \right]}{d k} \right\vert$ is greatest).  Furthermore we found that considering points up to $k \le k_{max}= 0.235 \,\rm{h \,Mpc^{-1}}$ yelds the optimal trade-off between accuracy on $\alpha$ and the $\chi^2/n_{ d.o.f.}$. So, in our analysis we do not include the four BOSS data with the highest momenta where BAO's are damped off.

The results for the $\alpha$ parameter from the monopole are listed in Table \ref{tab:results_mock}. The recovered values of $\alpha$ are in perfect agreement with $1$ for all redshift bins, which shows that our procedure provides an unbiased estimate of the BAO scale contained in the mock data. The constraints that we obtain range from the $1\,\%$ to the $1.8\,\%$ level.  Comparing with the analysis of the same data in \cite{Beutler:2016ixs},  we improved the accuracy on $\alpha$  on average by about 20\% for pre-reconstructed data while we obtain a comparable accuracy for the post reconstructed data.

{\color{red}
\begin{table}
\centering
\begin{tabular}{|l| l  c |  l c |}
\hline
 \multicolumn{5}{|c|}{\bf{Boss collaboration    \cite{Beutler:2016ixs}  }}\\[0.2cm]
 \hline
 \multicolumn{1}{|c|}{}& \multicolumn{2}{|c|}{Pre-reconstruction} & \multicolumn{2}{|c|}{Post-reconstruction}\\
 \hline
 & $\alpha$ & error & $\alpha$ & error \\
\hline
$0.2 < z < 0.5$  & 1.010 & 0.022 & 1.002 & 0.013 \\
$0.4 < z < 0.6$  & 1.010 & 0.019 & 1.003 & 0.012 \\
$0.5 < z < 0.75$ & 1.008 & 0.019 & 1.003 & 0.012 \\
\hline
\end{tabular}
\begin{tabular}{|l| l  c |  l c |}
\hline
 \multicolumn{5}{|c|}{\bf{Extractor procedure}}\\[0.2cm]
 \hline
 \multicolumn{1}{|c|}{}& \multicolumn{2}{|c|}{Pre-reconstruction} & \multicolumn{2}{|c|}{Post-reconstruction}\\
 \hline
 & $\alpha$ & error & $\alpha$ & error \\
\hline
$0.2 < z < 0.5$  & 1.005 & 0.018 & 1.001 & 0.012 \\
$0.4 < z < 0.6$  & 1.004 & 0.016 & 1.002 & 0.010 \\
$0.5 < z < 0.75$ & 1.003 & 0.015 & 1.002 & 0.011 \\
\hline
\end{tabular}
\caption{Results for the parameter $\alpha$ obtained by applying our analysis to mock galaxy data. The errors listed in this and in the following tables delimit the $68 \% \, {\rm C.L.}$ intervals.  
}
\label{tab:results_mock}
\end{table}

}



To gain a feeling on whether the error on $\alpha$ shown in Table \ref{tab:results_mock} is a credible one, we performed an extremely simple Fisher forecast, using the BOSS covariance matrix (including only the points in the $\left\{ \rho_i \right\}$ set),  and assuming a linear power spectrum with the only unknown parameter $\alpha$, namely 
\begin{equation}
P \left( k ,\, \alpha \right) = P^{nw,l} \left( k \right) +  P^{w,l} \left( \alpha k \right) \, {\rm e}^{- \Xi \, k^2} \;. 
\end{equation} 
The choice of the linear power spectrum is dictated by simplicity (since here we are only interested to determine if the error bar from our analysis is a credible one). Finally, the exponential damping accounts for the effect of the reconstruction. Specifically, we take $\Xi =0$ for the Fisher error corresponding to the post-reconstruction data (assuming in this idealized estimate that the reconstruction perfectly reproduces the oscillations in the initial linear data), while we take $\Xi^{1/2} =  6 \, h^{-1} \, {\rm Mpc}$ as obtained by inserting the linear power spectrum of the fiducial cosmology in eq. (\ref{xi-bao}). We performed this test for the lowest redhsift bin only, obtaining $\sigma_{\alpha,{\rm Fisher}} \simeq 0.0022$ for the post-reconstruction case, and  $\sigma_{\alpha,{\rm Fisher}} \simeq 0.0035$ for the pre-reconstruction case. Since $\alpha$ is the only varying parameter, the error obtained with this Fisher estimate must be smaller than that obtained in any realistic analysis, and therefore of those presented in Table  \ref{tab:results_mock}. We see that this is indeed the case. 

In order to test if our procedure provides an unbiased estimate of the $\alpha$ parameter, we repeated the analysis on the MultyDark Patchy mock catalogues but assuming a  model cosmology in eq.~\re{eq:model_BOSS} different from the one used in producing the mock catalogues. In this case, the obtained $\alpha$ parameter should compensate for the ``wrong'' $\alpha$ of the assumed cosmology. Following  eqs.~(7) and (8) of \cite{Gil-Marin:2015nqa}, we expect the extracted $\alpha$ from the fit, that we indicate as $\alpha_{exp}$, to be given by
\beq
\alpha_{exp} =\frac{\alpha^{MD}}{\alpha'} =\frac{r_s'}{r_s^{MD}}\left(\frac{H'(z)}{H^{MD}(z)}\right)^{1/3}\left(\frac{D^{MD}_A(z) }{D_A'(z)}\right)^{2/3}\,,
\label{alphafit}
\eeq
where $r_s$, $H(z)$, and $D_A(z)$ indicate the sound horizon at decoupling, and  the Hubble parameter and the angular diameter distance at the effective redshift of the measurement, for the MultyDark Patchy cosmology (MD) and for the assumed ``wrong'' cosmology (that we indicate with primes). We fix the latter as the ``QPM" model considered in \cite{Gil-Marin:2015nqa}, that we denote here as `Model 1', whose parameters are given in the third column of Table \ref{tab:cosmologies}. We also consider the ``Fiducial'' cosmology of  \cite{Gil-Marin:2015nqa}, that we denote here as `Model 2', whose parameters are given in the fourth column of Table \ref{tab:cosmologies}. The results are  shown in Tab.~\ref{tab:results_wrong_cosm}. It is clear that, even assuming a cosmology giving a 2 \% deviation in $\alpha$ with respect to the one used to generate the simulations, the correct scale can be recovered within the statistical error.

\begin{table}
\centering
\begin{tabular}{|c|c| cc|cc|}
\hline
 \multicolumn{4}{|c|}{\bf{Model 1}}\\ [0.2cm]
 \hline
 \multicolumn{1}{|c|}{}& \multicolumn{1}{c}{Expected} & \multicolumn{2}{|c|}{Extracted (POST)}\\
 \hline
 & $\alpha_{exp}$ &  $\alpha$ & error\\
\hline
$0.2 < z < 0.5$  & 1.023  &  1.028 & 0.012 \\
$0.4 < z < 0.6$  & 1.021  &  1.031 & 0.010 \\
$0.5 < z < 0.75$ & 1.020  & 1.029 & 0.011 \\
\hline
\end{tabular}
\begin{tabular}{|c|c| cc|cc|}
\hline
 \multicolumn{4}{|c|}{\bf{Model 2}}\\ [0.2cm]
 \hline
 \multicolumn{1}{|c|}{}& \multicolumn{1}{c}{Expected} & \multicolumn{2}{|c|}{Extracted (POST)}\\
 \hline
 & $\alpha_{exp}$ &  $\alpha$ & error \\
\hline
$0.2 < z < 0.5$  & 1.006  &  1.009 & 0.012 \\
$0.4 < z < 0.6$  & 1.007  &  1.009 & 0.011 \\
$0.5 < z < 0.75$ & 1.007  & 1.008 & 0.011 \\
\hline
\end{tabular}
\caption{Results for the $\alpha$ parameter extracted from the mock simulations (reconstructed) assuming a ``wrong'' cosmology, compared to the expected result computed from eq.~\re{alphafit} (first column).}
\label{tab:results_wrong_cosm}
\end{table}


\section{Application to BOSS DR12 data}\label{sdata}

Having tested our procedure, and found the optimal setup, with mock galaxies, we apply it to the experimental data.  We use the same pipeline as the one established in analyzing Mock data presented in  Section \ref{smock},  namely, we use the data \cite{Beutler:2016ixs} up to $k \le k_{max}= 0.225 \,\rm{h \,Mpc^{-1}}$ (respectively, up to $k \le k_{max}= 0.235 \,\rm{h \,Mpc^{-1}}$) for the pre-reconstruction (respectively, post-reconstruction) data). This corresponds to  $22$ (respectively, $23$) bins. This gives $44$ (respectively, $46$) data in each simultaneous NGC+SGC fit (for each redshift bin) that are analyzed with three free parameters ($A$, $b$, and $\alpha$), resulting in $44-3=41$ degrees of freedom (respectively, $46-3=43$ degrees of freedom) in each analysis.  

The BOSS analysis \cite{Beutler:2016ixs} uses all bins up to  $k / \left( h \, {\rm Mpc}^{-1} \right) \simeq 0.295$. This means $58$ data (in the combined NGC+SGC) analysis, which, accounting for the $10$ parameters that we have described at the end of Section \ref{sec:model}, corresponds to $48$ degrees of freedom. 

\begin{figure}
\centerline{
\includegraphics[width=0.4\textwidth]{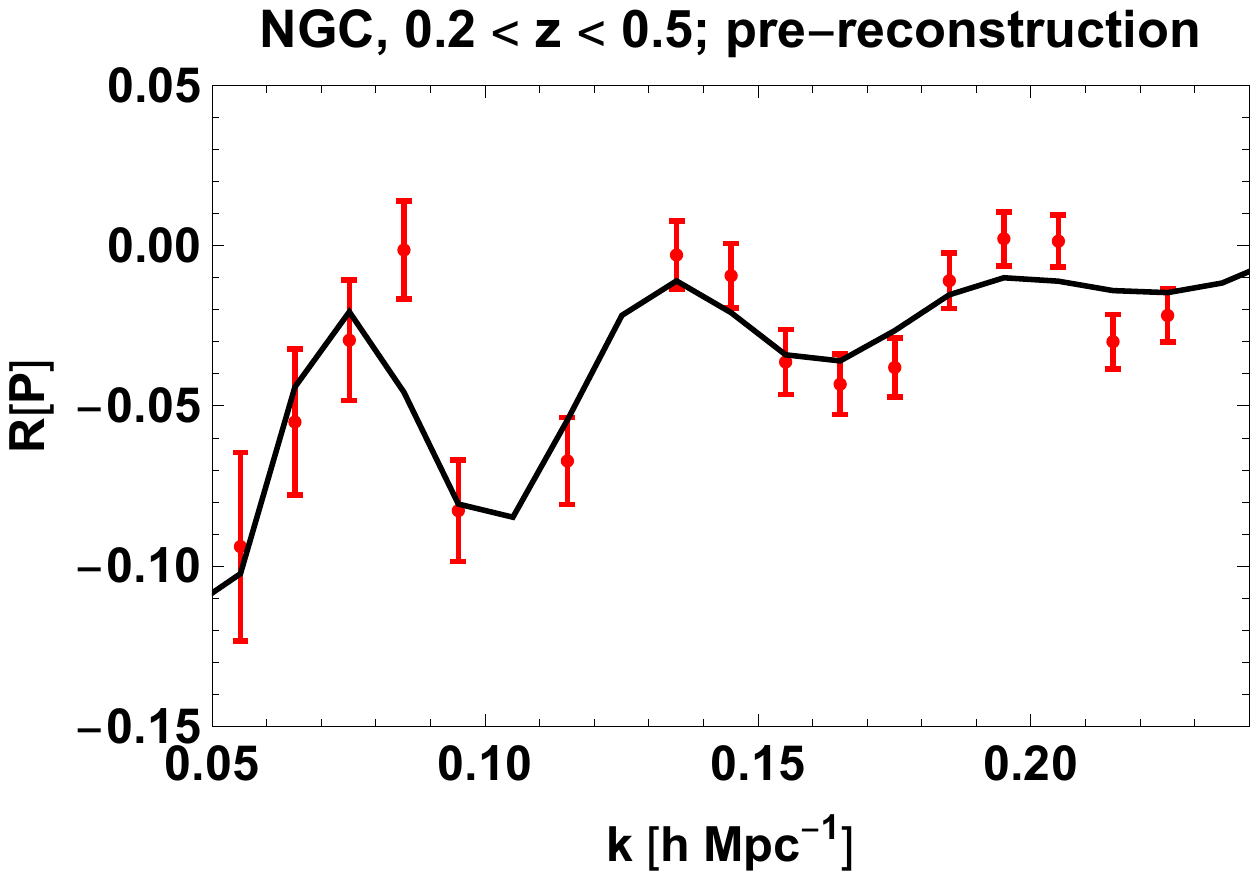} 
\includegraphics[width=0.4\textwidth]{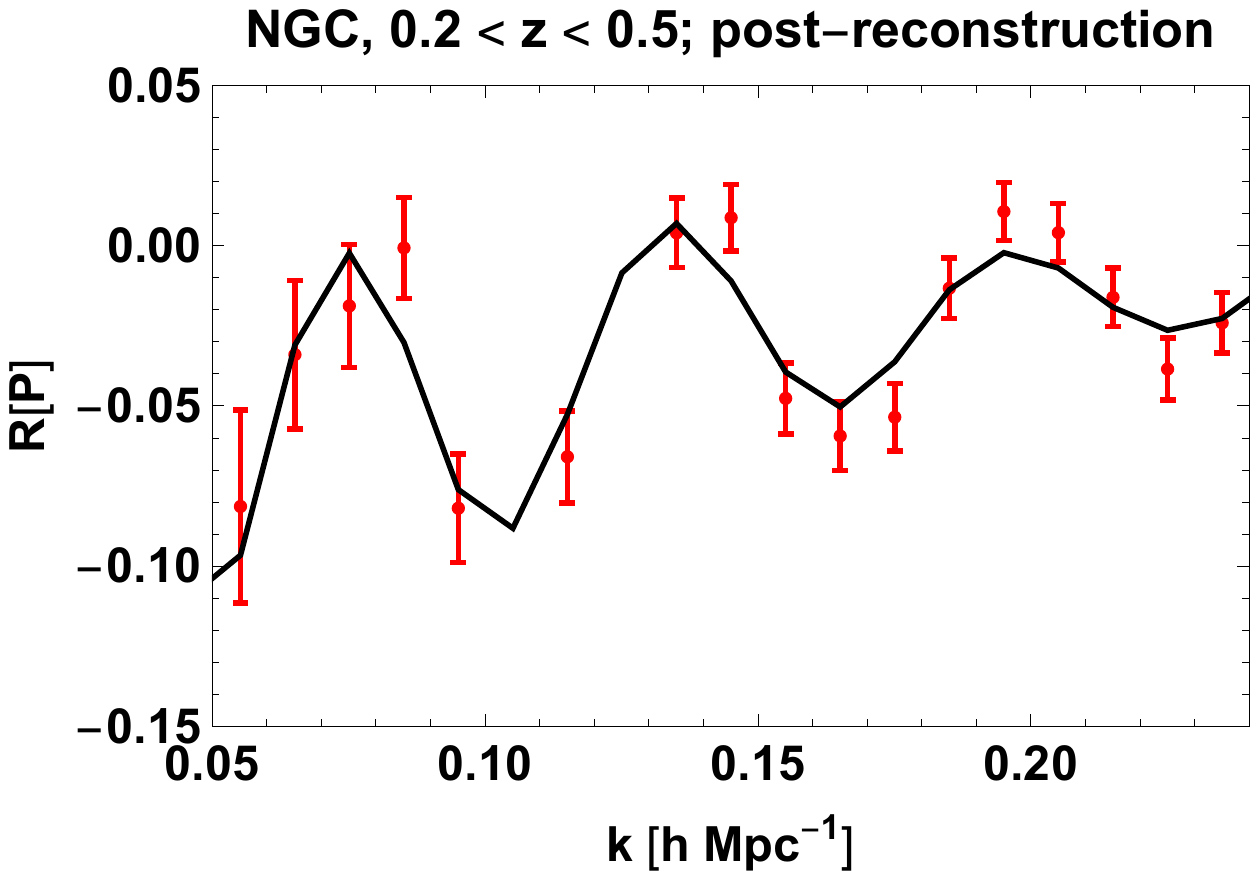} 
}
\centerline{
\includegraphics[width=0.4\textwidth]{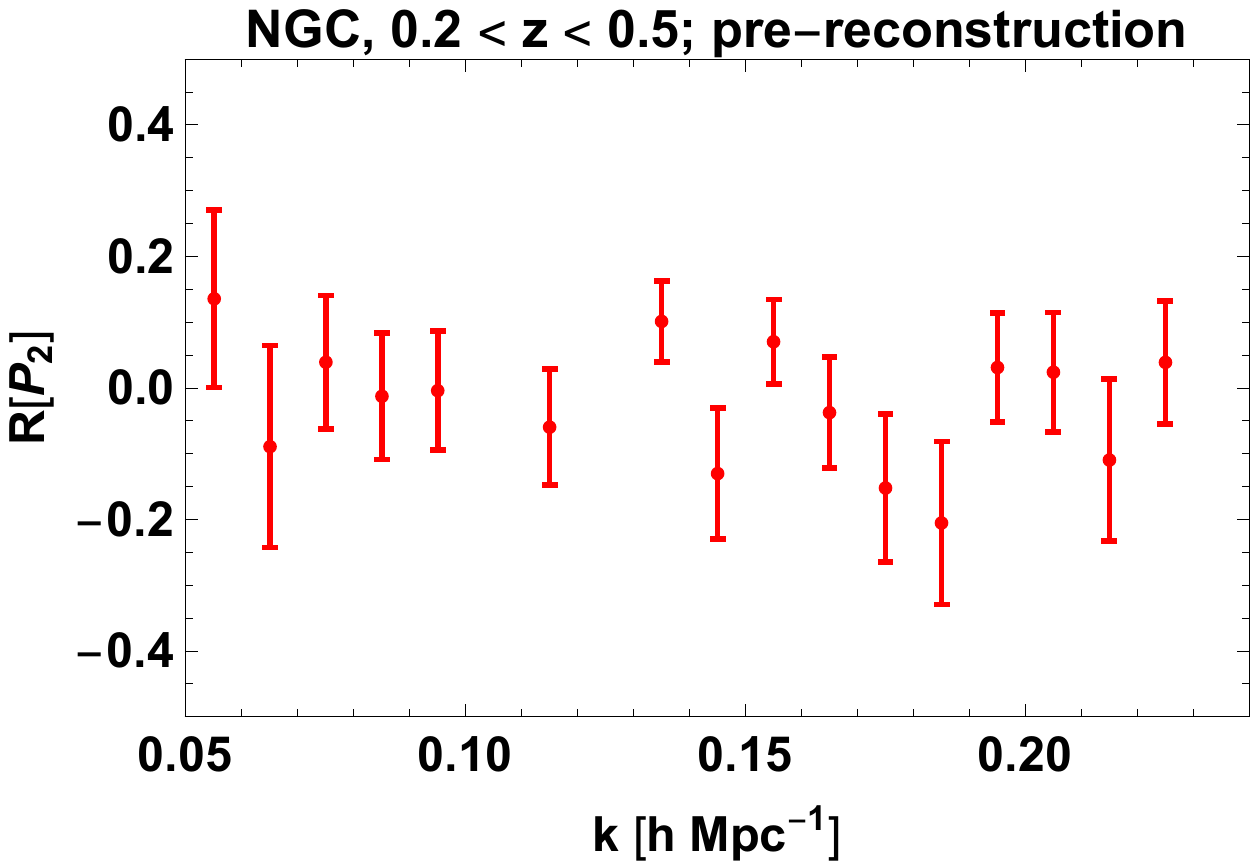} 
\includegraphics[width=0.4\textwidth]{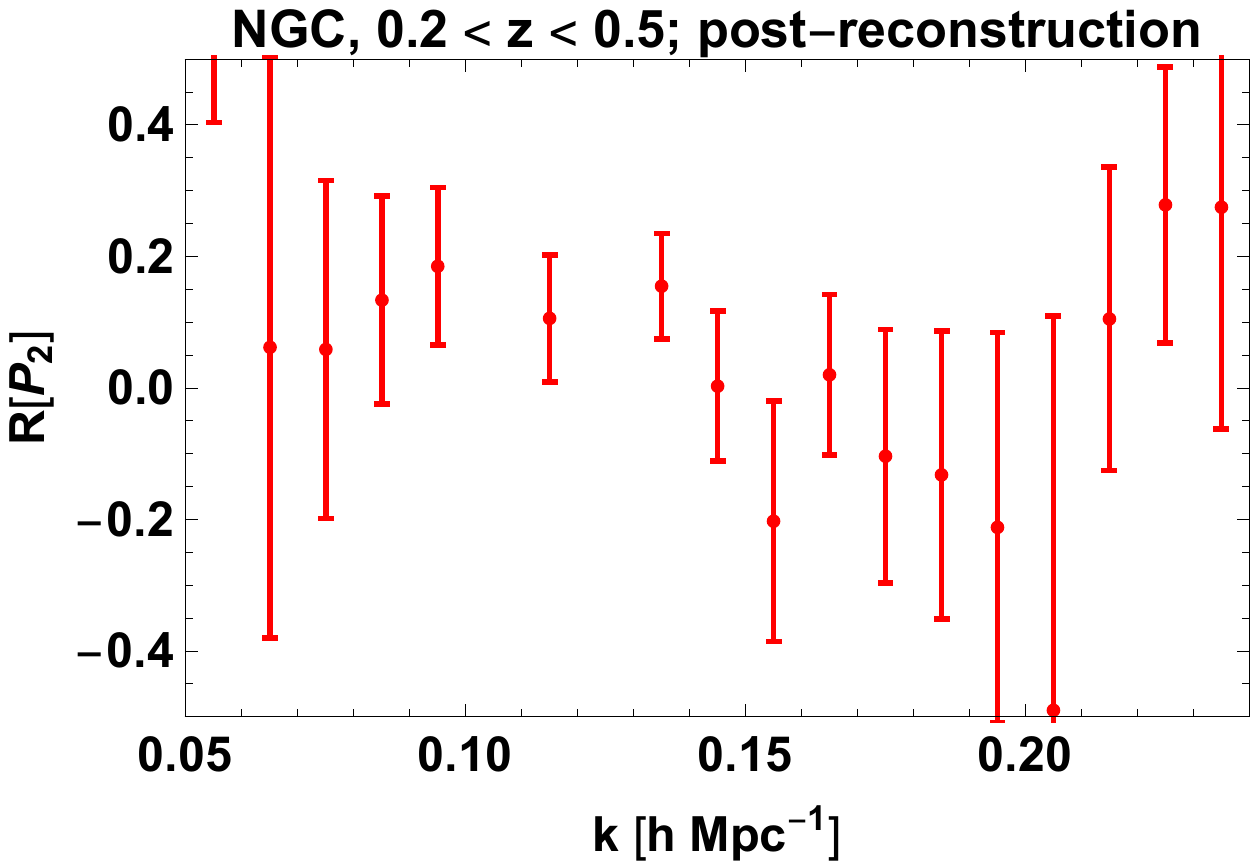}
}
\caption{First row: Extracted monopoles for the BOSS monopole data; we present the results for the 
pre-reconstruction (left panel) and post-reconstruction (right panel) cases for the lowest bin NGC data). 
The solid lines correspond to the best fit model of Eq.~\re{eq:model_BOSS}. Second row: Extracted monopoles for the BOSS quadrupole data. 
}
\label{Fig:extracted_monopole-dipole_BOSS}
\end{figure}

The reference cosmological model used in our analysis is the one given in the last column of Table \ref{tab:cosmologies}. The extracted monopoles are shown in the first row of Fig.  \ref{Fig:extracted_monopole-dipole_BOSS}, together with the data (we only show the data in the $\rho$ set, which, as explained above, are those used in the extractor procedure). We only show the lowest redshift bin for the NGC case, as the other cases present qualitatively similar features. By comparing the two panels of the first row, we can see the effect of reconstruction in enhancing the amplitude of the BAO wiggles. 

In the second row of Fig.  \ref{Fig:extracted_monopole-dipole_BOSS} we show instead the extractor applied to the quadrupole. The two panels show no evident oscillatory feature, meaning that the BAO signal is covered by the noise. In fact, adding the quadrupole to the analysis brings no improvement on the determination on $\alpha$. This is why the analysis of the BOSS data performed in this work has been restricted to the monopole.

\begin{figure}
\includegraphics[width=\textwidth]{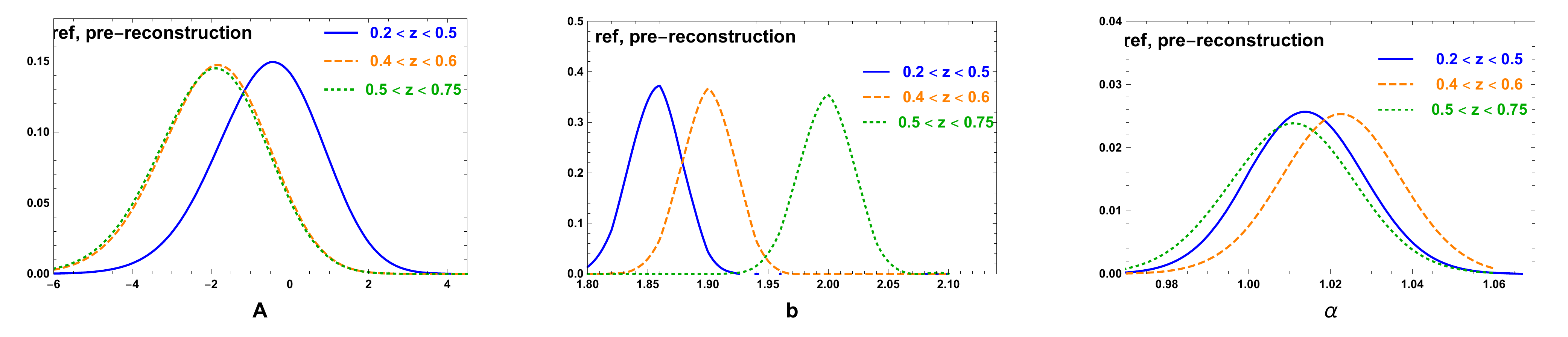} \\
\includegraphics[width=\textwidth]{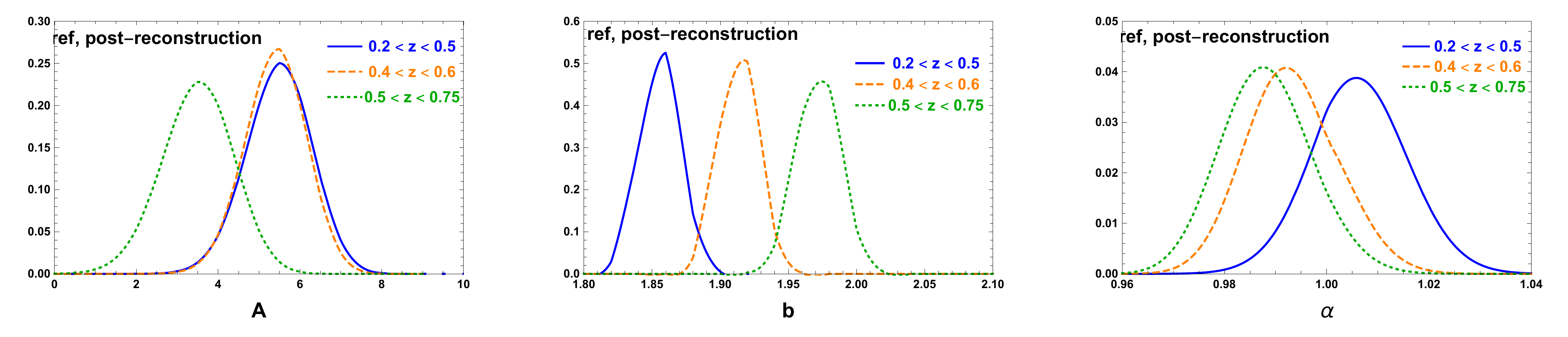} \\
\caption{Marginalized PDFs for the three parameters $A$, $b$, and $\alpha$, and for different redshift bins. 
The analysis refers to the BOSS DATA, confronted with the model (\ref{eq:model_BOSS}) for the reference cosmology (the last one in table \ref{tab:cosmologies}. The first (respectively, second) row refers to the  pre (respectrively, post) reconstruction data.}
\label{Fig:CHI2_ref}
\end{figure}

In Figure \ref{Fig:CHI2_ref} we show the marginalized PDFs for the comparison of the BOSS data with the  model of Eq.~\re{eq:model_BOSS}, and fo the reference cosmology.

Finally, the results for $\alpha$ are shown in Tab. \ref{tab:results_BOSS}.  As for the BOSS analysis, the accuracy on the dilation parameter $\alpha$  is increased with respect to the one extracted from Mock data. In \cite{Beutler:2016ixs}, it was explained as due to the fact that BAO features are reduced in the MultyDark Patchy mock catalogues.   In the case of pre-reconstructed data, our analysis improves the accuracy on the $\alpha$ parameter by about 20\%, while for the reconstructed ones the performances of the two analyses are essentially equal. From this we conclude that  there is no need to introduce additional nuisance parameters in our analysis (as those employed in the BOSS analysis). In fact, given the values for the $\chi^2$, this would result in an overfitting.

\section{Conclusions}\label{sconclu}
The Extractor procedure represents a promising approach to the study of BAOs, both from a theoretical and data analysis point of view. Despite using a simple model, Eq. \re{eq:model_BOSS}, the  procedure was able to reach a subpercent precision for all redshift bins, for reconstructed data, using only two nuisance parameter besides $\alpha$.  Applying the Extractor procedure on unreconstructed data gives constraints tighter by about 20\% than those obtained in the standard analysis. On the other hand, for reconstructed data, we get essentially the same accuracy as that reached in  \cite{Beutler:2016ixs}.

The BAO signature, as defined by the Extractor prescription, is mostly confined to the monopole PS. This is clear from Fig.~\ref{Fig:extracted_monopole-dipole_BOSS}, and from the analogous one for the  quadrupole from real data. Adding higher multipoles does not lead to any substantial improvement in our results.

There are a number of ways this method can be improved, also in view of future data. The exploration of the parameter space can go beyond the $\alpha$-parameterization,  by using Monte Carlo Markov Chain algorithms. On the theoretical side, the model we used used essentially a 1-loop approximation, while we showed that using the full TRG method presented in \cite{Noda:2017tfh}  improves the BAO scale extraction at virtually no extra computational cost. Although the modelling of RSD and halo bias presented here was very basic, the simple exponential  factor $\exp(-Ak^2)$ gave already good results. A more detailed description of these effect would likely improve further the constraints on $\alpha$. Moreover, by modelling the angular dependence of RSD more accurately, the quadrupole and, possibly, the hexadecapole data could be included in the analysis. We leave these improvements to future work.

\begin{table}
\centering
\begin{tabular}{|l| l  c c |  l c c |}
\hline
 \multicolumn{7}{|c|}{\bf{BOSS collaboration \cite{Beutler:2016ixs} }}\\ [0.2cm]
 \hline
 \multicolumn{1}{|c|}{}& \multicolumn{3}{|c|}{Pre-reconstruction} & \multicolumn{3}{|c|}{Post-reconstruction}\\
 \hline
 & $\alpha$ & error & $\chi^2 / {\rm gdl}$ & $\alpha$ & error & $\chi^2 / {\rm gdl}$ \\
\hline
$0.2 < z < 0.5$  & 1.006 & 0.016 & $48.5/48$ & 1.000 & 0.010 & 43.9/48 \\
$0.4 < z < 0.6$  & 1.016 & 0.017 & 64.8/48 & 0.9936 & 0.0082 & 32.8/48 \\
$0.5 < z < 0.75$ & 0.991 & 0.019 & 49.8/48 & 0.9887 & 0.0087 & 47.0/48 \\
\hline
 \multicolumn{7}{|c|}{\bf{Extractor procedure}}\\[0.2cm]
\hline
\multicolumn{1}{|c|}{}& \multicolumn{3}{|c|}{Pre-reconstruction} & \multicolumn{3}{|c|}{Post-reconstruction}\\
  \hline
 & $\alpha$ & error & $\chi^2 / {\rm gdl}$ & $\alpha$ & error & $\chi^2 / {\rm gdl}$ \\
\hline
$0.2 < z < 0.5$  & 1.014 & 0.014 & 32/41 &  1.006 & 0.009 & 31/43 \\
$0.4 < z < 0.6$  & 1.022 & 0.014 & 43/41 & 0.993 & 0.009 & 33/43 \\
$0.5 < z < 0.75$ & 1.011 & 0.015 & 30/41 & 0.988 & 0.009 & 40/43 \\
\hline
\end{tabular}
\caption{Results for the parameter $\alpha$ obtained with the standard procedure (BOSS collaboration, Ref. \cite{Beutler:2016ixs}) and by applying our analysis to BOSS experimental data. }
\label{tab:results_BOSS}
\end{table}

\section*{Acknowledgments}
It is a pleasure to thank Florian Beutler for providing us with the processed PS for the BOSS experimental data and mock galaxy catalogue, and for clarifying correspondence on the use of these data, and Takahiro Nishimichi for discussions and for collaborating at the early stages of the present analyses. We thank an anonymous referee for suggestions that have improved our analysis. 

MP acknowledges support from the European Union Horizon 2020 research and innovation programme under the Marie Sklodowska-Curie grant agreements Invisible- sPlus RISE No. 690575, Elusives ITN No. 674896 and Invisibles ITN No. 289442.

\section*{References}
\bibliographystyle{JHEP}
\bibliography{/Users/massimo/Bibliografia/mybib.bib}
\end{document}